\documentclass[twocolumn,reprint,showpacs,amsmath,amssymb]{revtex4}

\usepackage{graphicx}
\usepackage{dcolumn}
\usepackage{bm}
\usepackage{color}
\usepackage[hypertex]{hyperref}
\begin{document}

\title{Quasiparticle band structures and optical properties of strained monolayer
MoS$_{2}$ and WS$_{2}$}

\author{Hongliang Shi$^{1}$}
\author{Hui Pan$^{1}$}
\author{Yong-Wei Zhang$^{1}$}
\email{zhangyw@ihpc.a-star.edu.sg}
\author{Boris I. Yakobson$^{2}$}
\email{biy@rice.edu}
 \affiliation{$^1$Institute of High Performance Computing, A*STAR, Singapore 138632\\$^{2}$Department
 of Mechanical Engineering and Materials Science, and Department of Chemistry, Rice University,
 Houston Texas 77005}
\date{\today}

\begin{abstract}
{The quasiparticle (QP) band structures of both strainless and
strained monolayer MoS$_{2}$ are investigated using more accurate
many body perturbation \emph{GW} theory and maximally localized
Wannier functions (MLWFs) approach. By solving the Bethe-Salpeter
equation (BSE) including excitonic effects on top of the partially
self-consistent \emph{GW$_{0}$} (sc\emph{GW$_{0}$}) calculation, the
predicted optical gap magnitude is in a good agreement with available
experimental data. With increasing strain, the exciton binding energy
is nearly unchanged, while optical gap is reduced significantly. The
sc\emph{GW$_{0}$} and BSE calculations are also performed on
monolayer WS$_{2}$, similar characteristics are predicted and
WS$_{2}$ possesses the lightest effective mass at the same strain
among monolayers Mo(S,Se) and W(S,Se).
Our results also show that the electron effective mass decreases as the tensile
strain increases, resulting in an enhanced carrier mobility.
The present calculation results suggest a
viable route to tune the electronic properties of monolayer
transition-metal dichalcogenides (TMDs) using strain engineering for
potential applications in high performance electronic devices.}
\end{abstract}

\maketitle

\section{\label{sec:level1}INTRODUCTION}

Bulk TMDs consisting of two-dimensional (2D) sheets bonded to each
other through weak van der Waals forces have been studied
extensively owing to their potential applications in photocatalysis
\cite{r1} and catalysis \cite{r2,r3}. MoS$_{2}$, WS$_{2}$,
MoSe$_{2}$, and WSe$_2$ are examples of such TMDs. Recently, their
2D monolayer counterparts were successfully fabricated using
micromechanical cleavage method \cite{r4}. Since then, these
monolayer materials have attracted significant attention
\cite{r8,r9,r10,r12,r22,r23,r24,r26a}.

For monolayer MoS$_{2}$, a strong photoluminescence (PL) peak at
about 1.90 eV, together with peaks at about 1.90 and 2.05 eV of the
adsorption spectrum indicated that MoS$_{2}$ undergoes an indirect
to direct band gap transition when its bulk or multilayers form is
replaced by a monolayer \cite{r9,r10,r12}. Shifts of PL peak for the
monolayer MoS$_{2}$ were also observed experimentally, which was
attributed to the strain introduced by covered oxides \cite{r21}.
Theoretical studies employed density functional theory (DFT) method
also predicted monolayer MoS$_{2}$ to have a direct gap of 1.78 eV
\cite{r8}. It is known however that DFT does not describe excited
state of solids reliably. Furthermore, an important character in
low-dimensional systems is their strong exciton binding due to the
weak screening compared to bulk cases. Therefore, the good band gap
agreement between theoretical and experimental results for monolayer
MoS$_{2}$ may be a mere coincidence. As a channel material for
transistor application, theoretical simulations show that monolayer
WS$_{2}$ performs better than monolayer MoS$_{2}$ \cite{r38}. In
order to address above questions, it is important and necessary to
employ more accurate calculation method beyond DFT to investigate
the electronic structures of strained monolayer MoS$_{2}$ and
WS$_{2}$.

The most common method to circumvent drawback of DFT is the
\emph{GW} approximation \cite{r28}, in which, self-energy operator
$\Sigma$ contains all the electron-electron exchange and correlation
effects. The sc\emph{GW$_{0}$} approach, in which only the orbitals
and eigenvalues in \emph{G} are iterated, while \emph{W} is fixed to
the initial DFT \emph{W$_{0}$}, was shown to be more accurate in
many cases to predict band gaps of solids \cite{r30}. The
off-diagonal components of the self-energy $\Sigma$ should be
included in sc\emph{GW$_{0}$} calculations, since this inclusion has
been proved particularly useful for materials such as NiO and MnO
\cite{r31}. It is noted that $\Sigma$ within \emph{GW} approximation
is defined only on a uniform \emph{k} mesh in Brillouin-zone, due to
its non-locality. Therefore, unlike DFT band structure plot,  the QP
eigenvalues at arbitrary \emph{k} points along high symmetry lines
cannot be performed directly \cite{r32}. Started from the
sc\emph{GW$_{0}$} calculation, the QP band structure can be
interpolated using MLWFs approach. This combination was demonstrated
to be accurate and efficient for the sc\emph{GW} band structure
\cite{r32}. The \emph{GW} results were shown to agree well with the
photoemission data \cite{r32a}, while in order to reproduce the
experimental adsorption spectra, the consideration of attraction
between quasi-electron and quasi-hole (on top of GW approximation)
by solving BSE is indispensable \cite{r32a}, particularly for the
low-dimensional systems with strong excitonic effect. The main goal
of this study is to accurately predict the QP band structures and
optical spectra of monolayer MoS$_{2}$ as a function of strain by
adopting the DFT-sc\emph{GW$_{0}$}-BSE approach.

Strain in monolayer MoS$_{2}$ can be produced either by epitaxy on a
substrate or by mechanical loading.  It is well-known that strain
can be used to tune the electronic properties of materials. This is
particularly important for two-dimensional materials, which can
sustain a large tensile strain. In fact, shifts of PL peak observed
experimentally in monolayer MoS$_{2}$ was attributed to strain
\cite{r21}, and the magnetic properties of MoS$_{2}$ nanoribbons
could be tuned by applying strain \cite{r32b}.

By adopting the aforementioned approach, we systematically
investigate how the electronic structures and optical properties of
monolayer MoS$_{2}$ evolve as a function of strain. Our results show
that exciton binding energy is insensitive to the strain, while
optical band gap becomes smaller as strain increases. Based on the
more accurate band structures interpolated by MLWFs methods based on
sc\emph{GW$_{0}$} results, the effective masses of carriers are
calculated. In addition, this calculation approach is also employed
to investigate other monolayer TMDs, that is, WS$_{2}$, MoSe$_{2}$,
and WSe$_2$. Our results demonstrate that the effective mass
is decreased as the strain increases, and monolayer WS$_{2}$ possesses
the lightest carrier among the TMDs, suggesting that using monolayer
WS$_{2}$ as a channel material can enhance the carrier
mobility and improve the performance of transistor.

\section{\label{sec:level2}DETAILS OF CALCULATION}

Our DFT calculations were performed by adopting the generalized
gradient approximation (GGA) of PBE functional \cite{r33} for the
exchange correlation potential and the projector augmented wave
(PAW) \cite{r34} method as implemented in the Vienna ab initio
simulation package \cite{r35}. 12 valence electrons are included for
both Mo and W pseudopotentials. The electron wave function was
expanded in a plane wave basis set with an energy cutoff of 600 eV.
A vacuum slab more than 15 \AA\ (periodical length of c is 19 \AA)
is added in the direction normal to the nanosheet plane. For the
Brillouin zone integration, a 12$\times$12$\times$1 $\Gamma$
centered Monkhorst-Pack \emph{k}-point mesh is used. In the
following \emph{GW} QP calculations, both single-shot
\emph{G$_{0}$W$_{0}$} and more accurate sc\emph{GW$_{0}$}
calculations are performed. 184 empty conduction bands are included.
The energy cutoff for the response function is set to be 300 eV, the
obtained band gap value is almost identical to the case of 400 eV.
The convergence of our calculations has been checked carefully. For
the Wannier band structure interpolation, \emph{d} orbitals of Mo
(W) and \emph{p} orbitals of S (Se) are chosen for initial
projections. Our BSE spectrum calculations are carried out on top of
sc\emph{GW$_{0}$}. The six highest valence bands and the eight
lowest conduction bands were included as basis for the excitonic
state. BSE was solved using the Tamm-Dancoff approximation. Notice
that the applied strain in the present study is all equibiaxial,
unless stated otherwise.

\section{\label{sec:level3}RESULTS AND DISCUSSIONS}

\begin{figure*}
\includegraphics*[height=6.8cm,keepaspectratio]{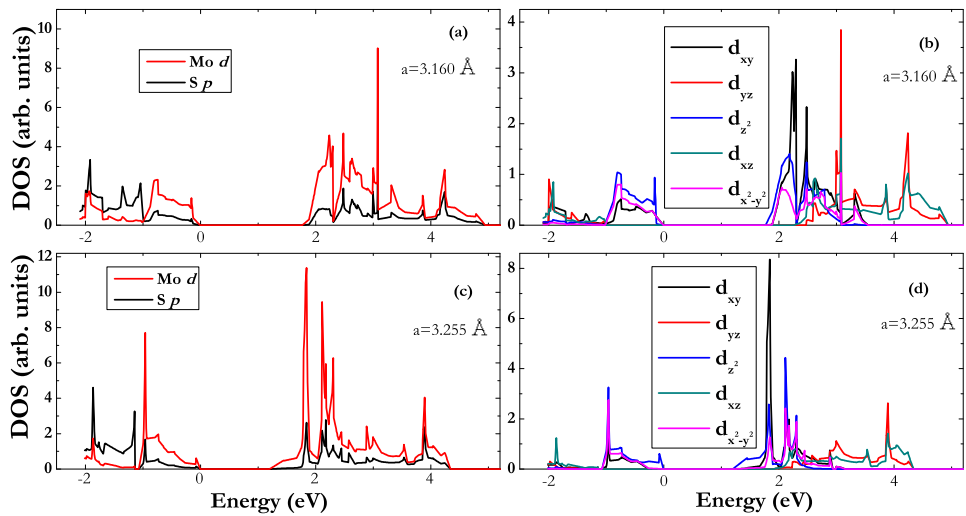}
\caption{\label{fig1}Projected density of states of \emph{d}
orbitals of Mo and \emph{p} orbitals of S (a and c) and decomposed
\emph{d} orbital of Mo (b and d) for monolayer MoS$_{2}$ at lattice
constants of 3.160 (a and b) and 3.255 \AA\ (c and d), respectively.
The latter corresponds to 3$\%$ tensile strain.}
\end{figure*}

\begin{figure}
\includegraphics*[height=5.6cm,keepaspectratio]{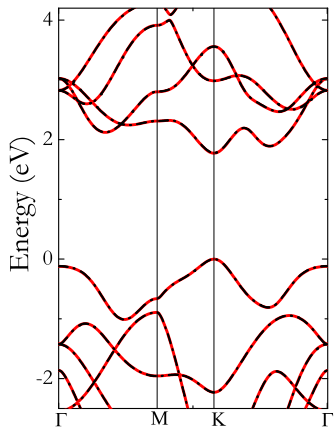}
\caption{\label{fig2}DFT band structures of monolayer MoS$_{2}$ at
lattice constant of 3.160 \AA. Red solid line: original band
structure obtained from a conventional first-principles calculation.
Black dash dot: Wannier-interpolated band structure. The Fermi level
is set to zero.}
\end{figure}

We first analyze the density of states (DOS) for monolayer
MoS$_{2}$. The \emph{d} orbitals of Mo and \emph{p} orbitals of S
contribute most to the states around the band gap, similar to
previous studies \cite{r22,r23,r24}. Fig. \ref{fig1} shows the
projected \emph{d} orbitals of Mo and \emph{p} orbitals of S as well
as the decomposed \emph{d} orbitals for monolayer MoS$_{2}$ at the
lattice of 3.160 \AA\ (the experimental lattice constant a of bulk
MoS$_{2}$ \cite{r22}) and under 3$\%$ tensile strain. Based on the
DOS, the \emph{d} orbitals of Mo and \emph{p} orbitals of S are
chosen as the initial projections in the Wannier interpolated
method. Fig. \ref{fig2} shows the identical DFT band structures of
monolayer MoS$_{2}$ obtained by the non-selfconsistent calculation
at fixed potential and Wannier interpolation method, respectively,
confirming that our choice of the initial projections and inner
window energy is appropriate. Based on the good results for
monolayer MoS$_{2}$, the same procedure is also employed for
remaining monolayer TMDs.

\begin{figure}
\includegraphics*[height=4.8cm,keepaspectratio]{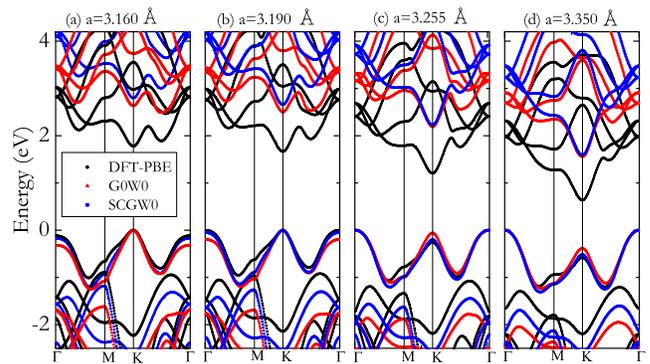}
\caption{\label{fig3}DFT, \emph{G$_{0}$W$_{0}$}, and
sc\emph{GW$_{0}$} QP band structures for monolayer MoS$_{2}$ at
lattice constants of (a) 3.160, (b) 3.190 (the optimized lattice
constant from this work), (c) 3.255, and (d) 3.350 \AA\,
corresponding to 0$\%$, 1$\%$, 3$\%$, and 6$\%$ tensile strain (with
reference to 3.160 \AA), respectively. The Fermi level is set to be
zero.}
\end{figure}

\subsection{\label{sec:level3}QP band structures of strained monolayer MoS$_{2}$}

The QP band structures of monolayer MoS$_{2}$ at four lattice
constants of 3.160, 3.190 (the optimized value from the present
work), 3.255, and 3.350 \AA\ are plotted in Fig. \ref{fig3},
corresponding to 0$\%$, 1$\%$, 3$\%$, and 6$\%$ tensile strains
(with reference to 3.160 \AA), respectively. As shown in Fig.
\ref{fig3}(a), the band structure obtained by DFT for strainless
MoS$_{2}$ is a direct band gap semiconductor with a band gap energy
of 1.78 eV, while the indirect band gap of 2.49 eV is predicted by
\emph{G$_{0}$W$_{0}$}. Obviously this \emph{G$_{0}$W$_{0}$} indirect
band gap is contrary to the PL observations \cite{r9,r10,r12}. The
QP band structures predicted by our sc\emph{GW$_{0}$} calculation
show that MoS$_{2}$ is a \emph{K} to \emph{K} direct band gap
semiconductor with a band gap energy of 2.80 eV. This prediction is
in excellent agreement with the recent calculation for MoS$_{2}$ at
the experimental lattice using full-potential linearized
muffin-tin-orbital method (FP-LMTO) \cite{r36}, which predicted a
\emph{K} to \emph{K} direct band gap of 2.76 eV.

It should be noted that in the 2D materials, the excitonic effect is
strong due to the weak screening. Thus it is important to consider
the attraction between the quasi-electron and quasi-hole by solving
the BSE discussed below in order to make the predicted optical gap
consistent with the optical spectra.

Fig. \ref{fig3}(b) shows the band structure of monolayer MoS$_{2}$
at 3.190 \AA\, corresponding to 1$\%$ strain. The DFT result
predicts the monolayer MoS$_{2}$ to be an indirect band gap with
\emph{K} to $\Gamma$ of 1.67 eV. Previous DFT studies also found
that monolayer MoS$_{2}$ already becomes an indirect semiconductor
under a  tensile strain of 1$\%$ \cite{r26a}. After \emph{GW}
correction, both of the \emph{G$_{0}$W$_{0}$} and sc\emph{GW$_{0}$}
QP band structures show that MoS$_{2}$ is still a direct
semiconductor with \emph{K} to \emph{K} band gaps of 2.50 and 2.66
eV, respectively. As the strain increases, shown in Fig.
\ref{fig3}(c) and \ref{fig3}(d), the DFT, \emph{G$_{0}$W$_{0}$} and
sc\emph{GW$_{0}$} all predict monolayer MoS$_{2}$ to be indirect.
The calculated indirect band gaps from  DFT, \emph{G$_{0}$W$_{0}$}
and sc\emph{GW$_{0}$} are 1.20 (0.63), 2.19 (1.56), and 2.23 (1.59)
for monolayer MoS$_{2}$ under strain of 3$\%$ (6$\%$), respectively.
As shown in Fig. \ref{fig3}, the value of band gap decreases as the
tensile strain increases, accompanying a shift of valence band
maximum (VBM) from \emph{K} to $\Gamma$ point and resulting in a
direct to indirect band gap transition, which was consistent well
with previous results \cite{r22,r26a}

\begin{figure}
\includegraphics*[height=5.8cm,keepaspectratio]{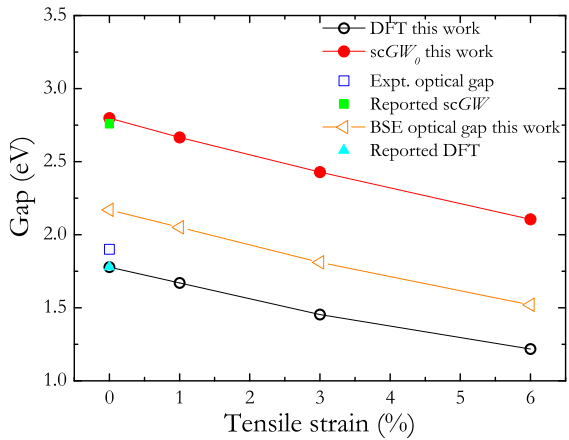}
\caption{\label{fig4}Band gaps for monolayer MoS$_{2}$ obtained by
DFT, sc\emph{GW$_{0}$}, and BSE. Reported experimental (Expt.)
\cite{r10}, DFT \cite{r26a}, and sc\emph{GW} \cite{r36} results are
also shown.}
\end{figure}

The \emph{K} to \emph{K} direct band gaps of monolayer MoS$_{2}$
obtained by DFT and sc\emph{GW$_{0}$} as a function of tensile
strain are plotted in Fig. \ref{fig4}. Clearly our DFT and
sc\emph{GW$_{0}$} results have the same trends, and accord well with
reported DFT \cite{r26a} (cyan triangle) and sc\emph{GW} \cite{r36}
(green solid square) results, respectively. Due to the more accurate
description of many body electron-electron interaction, the
sc\emph{GW$_{0}$} band gaps are enlarged about 1 eV compared to DFT
results. The optical gap shown in Fig. \ref{fig4} will be discussed
in the next subsection.

\subsection{\label{sec:level3}Excitonic effect in monolayer MoS$_{2}$}
\begin{figure}
\includegraphics*[height=10.6cm,keepaspectratio]{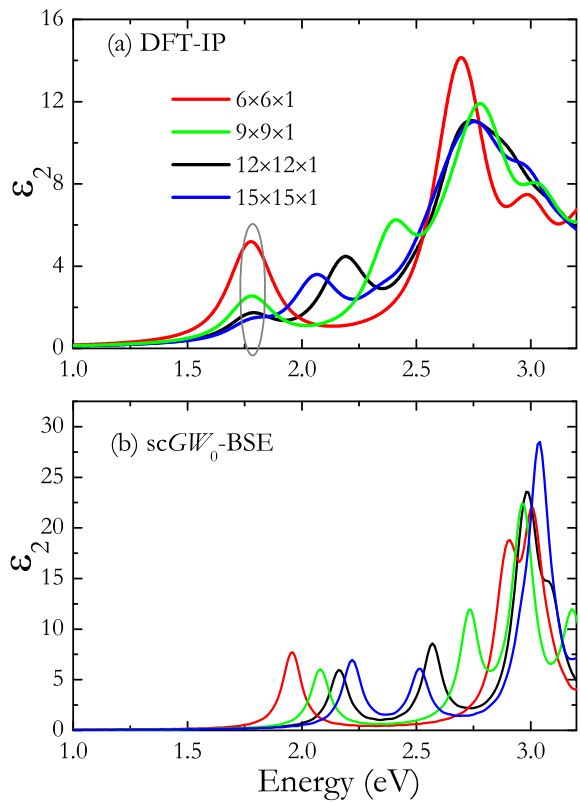}
\caption{\label{fig5}DFT-IP and sc\emph{GW$_{0}$}+BSE adsorption
spectra for monolayer MoS$_{2}$ at experimental lattice of 3.160
\AA\ (strainless case) obtained by different \emph{k}-point meshes.}
\end{figure}

In this subsection, the optical properties of monolayer MoS$_{2}$
are discussed in details. From the technical view, optical
transition simulation needs the integration over the irreducible
Brillouin zone using sufficiently dense \emph{k}-point mesh.
Naturally, the convergence of \emph{k}-point sampling is important.
First, for monolayer MoS$_{2}$ at strainless case (3.16 \AA), the
optical adsorption spectra $\varepsilon_{2}$
($\varepsilon_{xx}$=$\varepsilon_{yy}$) obtained by different
\emph{k}-point meshes are illustrated in Fig. \ref{fig5}(a), in
which the independent-particle (IP) picture is adopted within DFT
(DFT-IP) and no local filed effect is included at Hartree or DFT
level. The first peak at about 1.78 eV is observed clearly in all
the cases, corresponding to the \emph{K-K} direct transition. The
second significant peak located at about 2.75 eV is converged for
12$\times$12$\times$1 and 15$\times$15$\times$1 \emph{k}-point
meshes. Other peaks in adsorption spectra between the two
aforementioned dominated peaks mainly originate from different
irreducible \emph{k} points with unequal weights in different
\emph{k}-point meshes. According to our analysis of projected
density of states, the two significant peaks located in 1.78 and
2.75 eV correspond to \emph{d-d} and \emph{p-d} transitions,
respectively. Considering the dipolar selection rule only
transitions with the difference $\Delta$$l$ = $\pm$1 between the
angular momentum quantum numbers $l$ are allowed, \emph{i.e.}, the
atomic \emph{d-d} transition is forbidden. However, in the monolayer
MoS$_{2}$, due to the orbital hybridization, the VBM and conduction
band minimum (CBM) still have \emph{p} orbital contributions,
especially the former; thus the VBM to CBM transition dominated by
\emph{d-d} transition is still allowed. As expected, the strength of
this \emph{d-d} transition is weaker than the \emph{p-d} transition
as shown in Fig. \ref{fig5}(a).

\begin{table*}
\caption{\label{tab:table1}QP band gap, optical band gap and exciton
binding energy for monolayer MoS$_{2}$ and WS$_{2}$ are obtained by
QP sc\emph{GW$_{0}$} and BSE with and without spin-orbital coupling
(SOC) adopting different energy cutoffs and \emph{k}-point mesh. All
energies are in the unit of eV.}
\begin{ruledtabular}
\begin{tabular}{cccccccccccc}
& Energy cutoffs & \emph{k}-point & \emph{E}$_{g}$ &
\emph{E}$_{g}$(optical) &
Binding energy\\
           &&&&& \\
Monolayer MoS$_{2}$ (3.160 \AA)          & 400 and 200 &
6$\times$6$\times$1(SOC) & 2.89 & 1.87 & 1.01
          \\
          &                &6$\times$6$\times$1  & 2.99 & 1.96 & 1.03\\
          &                &9$\times$9$\times$1    & 2.84 & 2.08 & 0.76\\
          &                &12$\times$12$\times$1    & 2.78 & 2.16 & 0.62\\
          &                &15$\times$15$\times$1    & 2.76 & 2.22 & 0.54\\
                           & 600 and 300    &12$\times$12$\times$1  & 2.80 & 2.17 & 0.63\\
 &&&&& \\
Monolayer MoS$_{2}$ (3.190 \AA)          & 600 and 300    &12$\times$12$\times$1  & 2.66 & 2.04 & 0.62\\
 &&&&& \\
Monolayer WS$_{2}$ (3.155 \AA)          & 400 and 200   &6$\times$6$\times$1(SOC)  & 3.02 &1.97  &1.05 \\
          &                &6$\times$6$\times$1  & 3.28 & 2.21 &1.07  \\
          &                &9$\times$9$\times$1    & 3.12 & 2.34 & 0.78\\
          &                &12$\times$12$\times$1    & 3.06 & 2.43 & 0.63\\
          &                &15$\times$15$\times$1    & 3.05 & 2.51 &0.54 \\
                                        & 600 and 300    &12$\times$12$\times$1    & 3.11 & 2.46 & 0.65 \\
 &&&&& \\
Monolayer WS$_{2}$ (3.190 \AA)          & 600 and 300    &12$\times$12$\times$1    & 2.92 & 2.28 &0.64 \\
\end{tabular}
\end{ruledtabular}
\end{table*}

As for the BSE calculations, in order to reduce the computational
cost, we adopt 400 and 200 eV for the plane wave energy cutoff and
response function energy cutoff (short for 400 and 200 eV for energy
cutoffs), respectively, while the accuracy still can be guaranteed.
Taking the strainless monolayer MoS$_{2}$ for example, the
sc\emph{GW$_{0}$} band gap is 2.78 eV, resulting in only 0.02 eV
difference compared to 2.80 eV aforementioned using 600 and 300 eV
for energy cutoffs. The calculated BSE spectra for strainless
monolayer MoS$_{2}$ are plotted in Fig. \ref{fig5}(b). It is clearly
that as \emph{k}-point mesh refines, the first peaks have a
blueshift. For \emph{k}-point meshes 6$\times$6$\times$1,
9$\times$9$\times$1, 12$\times$12$\times$1, and
15$\times$15$\times$1, the sc\emph{GW$_{0}$} band gaps are 2.99,
2.84, 2.78, and 2.76 eV, respectively; the first adsorption peaks
(optical band gaps) are 1.96, 2.08, 2.16, and 2.22 eV.
Correspondingly, the exciton binding energies are 1.03, 0.76, 0.62,
and 0.54 eV, inferred from the difference between the QP
(sc\emph{GW$_{0}$}) and optical (sc\emph{GW$_{0}$}-BSE) gaps. These
calculated QP band gaps, optical gaps and exciton binding energies
are also listed in Table \ref{tab:table1}. The convergence trend is
obvious, particularly for the electronic band gap. However, due to
the limitation of computation resource, sc\emph{GW$_{0}$}
calculations with more dense \emph{k}-point mesh are not performed
here. Note that previous theoretical results showed a large value of
exciton binding energy for monolayer MoS$_{2}$. For example, a value
of 0.9 eV for monolayer MoS$_{2}$ (3.16 ${\rm{\AA}}$) was obtained
using empirical Mott-Wannier theory \cite{r36}; and a value of 1.03
eV was obtained by \emph{G$_{0}$W$_{0}$}-BSE calculations for
monolayer MoS$_{2}$ (3.18 ${\rm{\AA}}$) using 6$\times$6$\times$1
\emph{k}-point mesh and including spin-orbital coupling \cite{r36a},
which is the same as our above results using the same \emph{k}-point
mesh without spin-orbital coupling.

Experimentally, two close peaks observed in adsorption spectrum of
monolayer MoS$_{2}$ around 1.9 eV are due to the valence band
splitting caused by spin-orbital coupling. In our calculations, the
spin-orbital coupling is omitted unless otherwise stated and this
will not alter our main conclusions presented in the current study.
In order to make a comparison, we also performed the
sc\emph{GW$_{0}$}-BSE calculations with spin-orbital coupling using
6$\times$6$\times$1 \emph{k}-point mesh and 400 and 200 eV for
energy cutoffs. The two peaks in BSE adsorption spectrum locate at
1.87 and 2.05 eV and the corresponding exciton binding energy is
1.02 eV, consistent well with the aforementioned
\emph{G$_{0}$W$_{0}$}-BSE calculations using the same \emph{k}-point
mesh and energy cutoffs while different pseudopotentials
\cite{r36a}. Notice that the exciton binding energy obtained with
and without spin-orbital coupling for monolayer MoS$_{2}$ as shown
in Table \ref{tab:table1} is nearly the same, while the optical gap
in the former case shifts about 0.1 eV towards lower energy due to
the top valence band splitting of 0.17 eV according to our
sc\emph{GW$_{0}$} calculation.

For the evolution of exciton binding energy as a function of strain,
our results demonstrate that it is almost unchanged, i.e., 0.63 eV
(strainless), 0.62 eV (1\% strain), 0.62 eV (3\% strain), and 0.59
eV (6\% strain) (using 600 and 300 eV for energy cutoffs and
12$\times$12$\times$1 \emph{k}-point mesh). The direct optical gaps
are 2.17, 2.04, 1.81, 1.52 eV for the four cases shown in Fig.
\ref{fig3}, respectively, and also shown in Fig. \ref{fig4} using
the orange left triangles. The experimental optical gap for
monolayer MoS$_{2}$ was shown to be about 1.90 eV \cite{r10}. Since
there was no mention of specific lattice parameter, here it is
assumed to be the strainless case as shown in Fig. \ref{fig4}.
Notice that the consistency is good between our theoretical and
experimental results. If spin-orbital coupling is taken into
account, the consistency will be improved further, since the first
peak in the adsorption spectrum moves towards lower energy due to
the top valence band splitting. Most importantly, our results
demonstrate that the optical gap of monolayer MoS$_{2}$ is very
sensitive to tensile strain, which can be tuned by depositing
monolayer MoS$_{2}$ on different substrates \cite{r21}, whereas the
exciton binding energy is insensitive to it according to our current
results. This insensitivity is mainly because the hole and electron
are derived from the topmost valence and lowest conduction edge
states close to VBM and CBM that are significantly localized on Mo
sites (contributed by Mo \emph{d} orbitals) irrespective of the
magnitude of strain according to our DOS analysis.

We also notice that layer-layer distance or the length of vacuum
zone implemented in the periodical supercell methods has important
influence on the magnitude of the $G_{0}W_{0}$ band gap and the
exciton binding energy \cite{r37aa,r37a}. That is because the
long-range van der Waals forces originating from the nonlocal
electron-electron correlation is important in the layer TMDs. In
order to obtain an accurate exciton binding energy, the convergence
of \emph{k}-point mesh, the truncation of Coulomb interaction
\cite{r37aa}, and the resulting accurate QP band structure
($G_{0}W_{0}$ or sc\emph{GW}) are necessary. Compared to exciton
binding energy of 1.1 eV obtained by interpolation of $G_{0}W_{0}$
band gap \cite{r37a}, our exciton binding energy obtained using
denser \emph{k}-point is underestimated \cite{r37ab}, due to the
finite thickness of vacuum layer adopted in our periodical supercell
calculations. However, the magnitude of the optical gap is not
affected by the vacuum layer height according to our test (not shown
here). An interesting observation is that the optical gap of
monolayer MoS$_{2}$ is sensitive to the strain while the exciton
binding energy is not. Our results also show that the spin-orbital
coupling does not change the magnitude of exciton binding energy,
while the optical gap reduces towards experimental result due to the
band splitting at \emph{K} points and better consistency is
achieved.

\subsection{\label{sec:level3}Chemical bonding properties of monolayer MoS$_{2}$}

\begin{figure}
\includegraphics*[height=4.8cm,keepaspectratio]{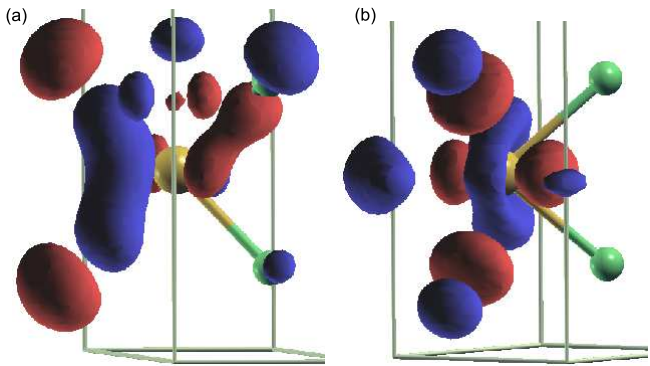}
\caption{\label{fig6}Isosurface plots of (a) valence-band and (b)
conduction band MLWFs for MoS$_{2}$ (at constant lattice of 3.16
\AA), at isosurface values $\pm$1.8/$\sqrt{\rm{V}}$, where V is the
unit cell volume, positive value red, and negative value blue. (a)
is Mo \emph{d}$_{x^2-y^2}$-like function showing bonding with the S
\emph{p}$_{x}$(\emph{p}$_{y}$) orbital, and (b) Mo
\emph{d}$_{z^2}$-like function showing antibonding with the S
\emph{p}$_{x}$(\emph{p}$_{y}$) orbital.}
\end{figure}

In order to gain further insight into the electronic structures
under the tensile strain, we revisit the DOS shown in Fig.
\ref{fig1}. For the strainless case, the VBM states at \emph{K}
mainly originate from Mo (\emph{d}$_{xy}$ + \emph{d}$_{x^2-y^2}$),
and S (\emph{p}$_{x}$ + \emph{p}$_{y}$) (decomposed \emph{p}
orbitals not shown in Fig. \ref{fig1}). The CBM at \emph{K} is
mainly contributed by Mo \emph{d}$_{z^2}$ and S (\emph{p}$_{x}$ +
\emph{p}$_{y}$). The Mo \emph{d} and S \emph{p} orbitals hybridize
significantly, therefore Mo and S form covalent bond. MLWFs can also
illustrate the chemical bonding properties of solids \cite{r37}. The
MLWFs shown in Fig. \ref{fig6} were constructed in two groups. The
first group was generated from \emph{d} guiding functions on Mo. The
energy window contains the topmost valence bands. Isosurface plots
of the Mo \emph{d}$_{x^2-y^2}$ MLWFs shown in Fig. \ref{fig6}(a)
show \emph{d}$_{x^2-y^2}$ orbitals form covalent bonding with
\emph{p}$_{x}$(\emph{p}$_{y}$) orbitals. The second group that MLWFs
for the low-lying conduction bands were also generated from Mo
\emph{d} guiding functions. Isosurface plots of the Mo
\emph{d}$_{z^2}$ MLWFs shown in Fig. \ref{fig6}(a) show
\emph{d}$_{z^2}$ oritals form antibonding with
\emph{p}$_{x}$(\emph{p}$_{y}$) orbitals. The chemical bonding
characters demonstrated by MLWFs are consistent well with our DOS
analysis shown in Fig. \ref{fig1}.

\subsection{\label{sec:level3}QP band structures and optical properties of strained monolayer WS$_{2}$}

\begin{figure}
\includegraphics*[height=4.8cm,keepaspectratio]{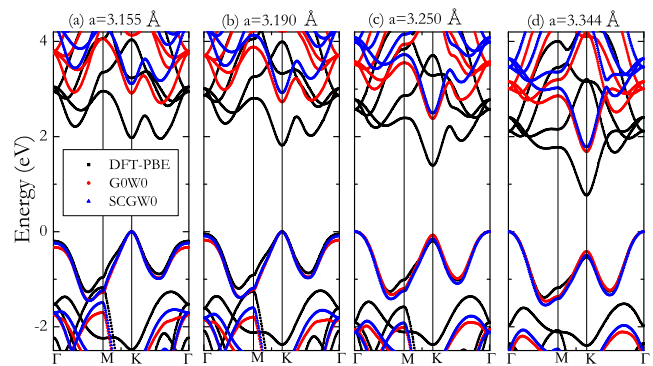}
\caption{\label{fig7}DFT, \emph{G$_{0}$W$_{0}$}, and
sc\emph{GW$_{0}$} QP band structures for WS$_{2}$ at lattice
constants of (a) 3.155, (b) 3.190 (optimized lattice constant this
work), (c) 3.250, and (d) 3.344 \AA\,, corresponding to 0$\%$,
1$\%$, 3$\%$, and 6$\%$ tensile strain (with reference to 3.155
\AA), respectively. The Fermi level is set to be zero.}
\end{figure}

The QP band structures of monolayer WS$_{2}$ under tensile strain
are also investigated, motivated by its better performance than
monolayer MoS$_{2}$ used as a channel in transistor devices
\cite{r38}. The calculation results are illustrated in Fig.
\ref{fig7}. Similar to monolayer MoS$_{2}$, the sc\emph{GW$_{0}$} QP
band structures of monolayer WS$_{2}$ also undergo a direct to
indirect band gap transition as tensile strain increases. The direct
band gaps for the strainless (at the experimental lattice of 3.155
\AA\ \cite{r22}) and under 1$\%$ tensile strain cases are 3.11 and
2.92 eV, respectively, and the latter corresponds to the optimized
lattice constant for monolayer WS$_{2}$ from this work. The
corresponding indirect band gaps under 3$\%$ and 6$\%$ tensile
strains are 2.49 and 1.78 eV, respectively. Note that for the
strainless case, our DFT result predicts monolayer WS$_{2}$ to be an
indirect band gap semiconductor with CBM only about 16 meV lower
than the lowest conduction band at \emph{K} points, which is
contrary to recent full potential methods \cite{r22}. The difference
may be originated from the technical aspect of these calculations,
such as the employed pseudopotential method \cite{r39}. However,
after the \emph{GW} correction, a correct direct band gap is
achieved.

For optical properties of monolayer WS$_{2}$, our calculated QP band
gaps, optical gaps and exciton binding energies are also listed in
Table \ref{tab:table1}. It is obvious that the monolayer WS$_{2}$
presents many similar properties compared to monolayer MoS$_{2}$,
for example, the gaps and exciton binding energy also demonstrate a
convergence trend as \emph{k}-point mesh increases; the spin-orbital
coupling has little influence on the magnitude of the exciton
binding energy. Notice that our sc\emph{GW$_{0}$} calculation
predicts the top valence band splitting of monolayer WS$_{2}$ to be
0.44 eV, larger than that of monolayer MoS$_{2}$ of 0.17 eV, because
W is much heavier than Mo. The resulting first peak in BSE
adsorption spectrum shifts 0.24 eV towards lower energy, also larger
that that of monolayer MoS$_{2}$ of 0.1 eV correspondingly. As for
the strain effect, the BSE optical gap at our optimized lattice
constant of 3.190 \AA\ is 2.28 eV, while at 3.16 \AA\, it is 2.46
eV, as shown in Table \ref{tab:table1}. The former corresponding to
1\% tensile strain, results in 0.18 eV reduction of band gaps. This
demonstrates that the band gaps and optical gaps are also very
sensitive to tensile strain, whereas the exciton binding energy is
not. Based on above analysis, we predict the exciton binding energy
of monolayer WS$_{2}$ is similar to that of MoS$_{2}$.
Experimentally, the PL maximum of monolayer WS$_{2}$ locates between
1.94 and 1.99 eV \cite{r40}. Considering the large shift of the peak
in the BSE adsorption spectrum caused by spin-orbital coupling, our
results at optimized lattice of 3.190 \AA\ are consistent well with
above experimental result \cite{r40}.

According to our above sc\emph{GW$_{0}$} and BSE calculations for
monolayer MoS$_{2}$ and WS$_{2}$, it is clear that the self energy
within the sc\emph{GW$_{0}$} calculations enlarges the band gap by
accounting for the many body electron-electron interactions more
accurately, while the strong excitonic effect results in a
significant reduction of the band gap. Combining the two opposite
effects on band gaps, the final resulting optical gap is consistent
well with DFT band gaps. Therefore, the good band gap agreement
between DFT and experiment is only a coincidence due to the fact
that QP band gap correction is almost offset by exciton binding
energy. This phenomenon was also observed in monolayer of hybridized
graphene and hexagonal boron nitride, which also have strong
excitonic effect \cite{r42}.

We also perform the sc\emph{GW$_{0}$} QP band structures for
monolayer MoS$_{2}$ and WS$_{2}$ under 1$\%$ compressive strains.
Our results show that the compressed MoS$_{2}$ has a direct band gap
of 2.97 eV, while the compressed WS$_{2}$ has an indirect band gap
of 3.13 eV and \emph{K} to \emph{K} direct gap of 3.30 eV.

Our sc\emph{GW$_{0}$} results show that both MoSe$_{2}$ and
WSe$_{2}$ are also a direct semiconductor at the strainless state.
The experimental lattice constants \cite{r22} for MoSe$_{2}$ and
WSe$_{2}$ are 3.299 and 3.286 \AA\ and the optimized lattice
constants are 3.327 and 3.326 \AA, respectively; their direct
\emph{K-K} band gaps are 2.40 and 2.68 eV at experimental lattices
and 2.30 and 2.50 eV at the optimized lattices. Compared to the
experimental lattice, the optimized lattice corresponds to 0.86$\%$
(1.22$\%$) tensile strain for MoSe$_{2}$ (WSe$_{2}$), and the band
gap also decreases with increasing tensile strain.

\begin{table*}
\caption{\label{tab:table2}Electron and hole effective masses
($m^*$) derived from partially sc\emph{GW$_{0}$} QP band structures
for monolayer MoS$_{2}$, WS$_{2}$, MoSe$_{2}$, and WSe$_{2}$ at
different strains. The effective masses at $K$ and $\Gamma$ points
are along \emph{K}$\Gamma$ and \emph{M}$\Gamma$ directions,
respectively.}

\begin{ruledtabular}
\begin{tabular}{cccccccccc}
   & &Compressive (1$\%$) &Experimental lattices & Optimized lattices & Tensile (3$\%$)&Tensile (6$\%$) & &\\
\hline \\
  MoS$_{2}$  &K$_{e}$ &0.40  &0.36 (0.35$^{a}$,0.60$^{b}$) &0.32  &0.29        &0.27\\
             &K$_{h}$ &0.40  &0.39 (0.44$^{a}$,0.54$^{b}$) &0.37                     \\
             &$\Gamma_{h}$&&&              &1.36        &0.90\\

  WS$_{2}$   &K$_{e}$ &  &0.27 &0.24  &0.22        &0.20\\
             &K$_{h}$ &  &0.32 &0.31                     \\
             &$\Gamma_{h}$&&&              &1.24       &0.79\\

 MoSe$_{2}$  &K$_{e}$ &  &0.38 &0.36  &        &\\
             &K$_{h}$ &  &0.44 &0.42                     \\

 WSe$_{2}$  &K$_{e}$ &  &0.29 &0.26  &        &\\
             &K$_{h}$ &  &0.34 &0.33                     \\
\end{tabular}
$^a$Effective masses listed here are averages of the longitudinal
and transverse values in Ref. \cite{r36}.\\ $^b$ Effective masses
listed here are averages of the curvatures along the $\Gamma$$K$ and
$KM$ directions in Ref. \cite{r36a}
\end{ruledtabular}
\end{table*}

\subsection{\label{sec:level3}Effective mass}
Based on the more accurate sc\emph{GW$_{0}$} QP band structures, the
effective mass of carriers for TMDs are calculated by fitting the
bands to a parabola according to $E=\frac{\hbar^2k^2}{2m_{e}m^*}$,
where $m_{e}$ is the electron static mass. A \emph{k}-point spacing
smaller than 0.03 \AA$^{-1}$ is used to keep parabolic effects.
Electron and hole effective masses ($m^*$) at different strains are
collected in Table \ref{tab:table2}. For MoS$_{2}$ under different
strains, the CBM always locates in \emph{K} point, and the electron
effective mass $K_{e}$ increases with increasing compressive strain
while decreases with increasing tensile strain. As for hole,
initially the effective mass also decreases as the tensile strain
increases. After the direct to indirect gap transition, VBM shifts
to $\Gamma$ with heavier hole, which also decreases as the tensile
strain increases. Compared to the effective masses of 0.64 and 0.48
for hole and electron at \emph{K} point based on DFT calculation
performed at the experimental lattice \cite{r22} for MoS$_{2}$, the
effective masses are reduced due to the \emph{GW} correction in our
study.

It is noted that the carrier effective masses obtained by our
sc\emph{GW$_{0}$} calculations do not include the spin-orbital
coupling effect. Compared with those including spin-orbital effect
for monolayer MoS$_{2}$ \cite{r36}, it is found that the electron
effective masses are in good agreement while the present hole
effective mass is slightly smaller. This is mainly because the
spin-orbital coupling alters the curvature of the topmost valence
band close to VBM while the lowest conduction band close to CBM is
not affected. The large difference between sc\emph{GW}
(sc\emph{GW$_{0}$}) and \emph{G$_{0}$W$_{0}$} result \cite{r36a} may
be due to the poor \emph{k}-points sampling and non-self-consistent
(one-shot) GW calculations of the latter.

For WS$_{2}$, MoSe$_{2}$, and WSe$_{2}$, their masses also show
similar behaviors. It is noted that at the same strain level, the
electron effective mass of WS$_{2}$ is the lightest; and electron
effective mass decreases as strain increases, making WS$_{2}$ more
attractive for high performance electronic device applications since
a lighter electron effective mass can lead to a higher mobility.
Theoretical device simulations also demonstrated that as a channel
material, the performance of WS$_{2}$ is superior to that of other
TMDs \cite{r38}.

\section{\label{sec:level4}SUMMARY}

In summary, the QP band structures of monolayer MoS$_{2}$ and
WS$_{2}$ at both strainless and strained states have been studied
systematically. The sc\emph{GW$_{0}$} calculations are found to be
reliable for such calculations.  Using this approach, we find they
share many similar behaviors. For the optical properties of
monolayer MoS$_{2}$, exciton binding energy is found to be
insensitive to the strain. Our calculated optical band gap is also
consistent with experimental results. In addition, we find that the
electron effective masses of monolayer MoS$_{2}$, WS$_{2}$,
MoSe$_{2}$, and WSe$_{2}$ decrease as the tensile strain increases,
and WS$_{2}$ possesses the lightest mass among the four monolayer
materials at the same strain. Importantly, the present work
highlights a possible avenue to tune the electronic properties of
monolayer TMDs using strain engineering for potential applications
in high performance electronic devices.

\section{\label{sec:level4}ACKNOWLEDGMENTS}

Work at Rice was supported by the Office of Naval Research MURI
project and by the Robert Welch Foundation (C-1590).

\end{document}